\documentclass[11pt]{article}
\usepackage{graphicx}
\usepackage{epsfig}
\usepackage{graphicx,colordvi}
\usepackage[dvips]{color}
\usepackage{amssymb}
\usepackage{amsmath,amsfonts}
\usepackage{url}
\usepackage{bm}
\usepackage{longtable}
\def\t{\mbox{tr}}

\def\cG0{{\cal G}_0}

\def\cG{{\cal G}}

\def \t2g{t$_{2g}$}

\begin{document}
\parskip 1ex

\setcounter{section}{0}
\setcounter{figure}{0}

\begin{center}

Electronic Polarons, Cumulants and Doubly Dynamical Mean Field Theory:
Theoretical Spectroscopy for Correlated and Less Correlated Materials

Silke Biermann and Ambroise van Roekeghem

Centre de Physique Th\'eorique, Ecole Polytechnique, 
CNRS, Universit\'e Paris Saclay, 91128
Palaiseau, France

\end{center}

\begin{abstract}
The use of effective local Coulomb interactions that are
{\it dynamical}, that is, frequency-dependent, is an efficient
tool to describe the effect of long-range Coulomb interactions
and screening thereof in solids. The dynamical character
of the interaction introduces the coupling to screening degrees
of freedom such as plasmons or particle-hole excitations
into the many-body description.
We summarize recent progress using these concepts, putting
emphasis on dynamical mean field theory (DMFT) calculations 
with dynamical interactions (``doubly dynamical mean field 
theory'').
We discuss the relation to the combined GW+DMFT method and
its simplified version ``Screened Exchange DMFT'', as well
as the cumulant schemes of many-body perturbation theory.
On the example of the simple transition metal SrVO$_3$,
we illustrate the mechanism of the appearance of plasmonic
satellite structures in the spectral properties, and discuss
implications for the low-energy electronic structure.
\end{abstract}

\section{Theoretical Spectroscopy: from many-body perturbation
theory to dynamical Hubbard interactions}

Determining the behavior of a single electron in a periodic
potential, created for example by the ions in a cristalline
solid, is a textbook exercise of quantum mechanics.
Determining the wave function of {\it all} the electrons in
the solid, however, is an intractable many-body problem.
The Pauli principle imposes full antisymmetry under exchange
of any two electrons to this object, and electronic Coulomb
interactions prevent it from being a simple Slater determinant.

The good news is that in practice the knowledge of the full
many-body wave function of the inhomogeneous electron gas in the
solid is barely necessary: the relevant electronic properties
are determined by the low-energy response to external 
perturbations, and the knowledge of these low-energy excitations 
requires much less information than the full ground-state wave 
function.
In this sense, solid state spectroscopies are a most efficient
means for characterizing the properties of a solid state system.
An important example are photoemission experiments -- angle-resolved
or angle-integrated -- where information about the electron removal
and addition spectra are obtained.
Within the simplest possible model for the photoemission process,
the so-called ``three-step model'', the photocurrent can be 
expressed in terms of the one-particle spectral function
$A(k, \omega) = - \frac{1}{\pi} Tr \Im G(k, \omega)$, and
computing this quantity from first principles, that is,
without adjustable parameters, is one of 
the central challenges of modern theoretical spectroscopy.

Important progress has been achieved over the last decades 
within many-body perturbation theory: a first order expansion 
of the many-body self-energy $\Sigma$ in the screened Coulomb 
interaction $W$ \cite{Hedin-1965,Hedin_book}
leads to a conceptually simple approximation
$\Sigma = i G W$ which can be calculated within realistic
electronic structure codes based on density functional theory (DFT).
For reviews of successful applications of the GW approximation
and developments based on it, we refer the reader to 
\cite{ferdi, onida}. 
For more strongly correlated materials, where perturbative
techniques reach their limits, the last 15 years have seen
the development of a non-perturbative theory, combining 
dynamical mean field theory (DMFT) \cite{georges}
with density functional
theory. This so-called ``DFT+DMFT'' approach 
\cite{Anisimov97, Lichtenstein98} builds on the
success of DMFT for the description of lattice models for
correlated fermions but extends its scope to the solid by
treating a realistic (multi-orbital) Hamiltonian with effective
local Coulomb interaction, often parametrized as Hubbard 
$U$ and Hund's $J$. 

While DFT+DMFT -- at least in its early implementations --
can simply be understood as the DMFT solution of a multi-orbital
Hubbard model (for reviews see \cite{biermann_ldadmft, vollkot, held-psik},
for some more modern implementations \cite{aichhorn09, lechermann}) 
recent efforts have been spent in order
to promote DMFT-based techniques to truly first-principles
techniques \cite{biermann-JPCM}.
This implies not only addressing the question of
how to relate effective local Hubbard interactions to the
full Coulomb interactions in the continuum (while taking
care to avoid double-counting of screening), that is the
{\it ab initio} calculation of the infamous effective local
``Hubbard U''; since at the
DFT level no rigorous distinction between contributions of
``correlated degrees of freedom'' and ``uncorrelated'' ones
can be made, a truly double-counting free theory can only
be achieved by eliminating the reference to the DFT Kohn-Sham
Hamiltonian altogether. A successful route is the combination of Hedin's
GW approximation with DMFT, the so-called GW+DMFT method 
\cite{gwdmft, gwdmft_proc2, gwdmft_proc1, PhysRevLett.92.196402}.
A summary of recent progress along these lines can be found
in \cite{biermann-JPCM}; for most recent applications both,
in the model and realistic electronic structure context we
refer the reader to Refs. \cite{ayral, ayral-prb, 
jmt_svo, jmt_svo2, hansmann}.
The common point between the GW method and the combined 
GW+DMFT scheme is the absence of adjustable interaction
parameters. Both theories can be viewed as approximations
to a free energy functional \cite{Almbladh_1999}, 
where the free energy of the
solid is written as a functional of the Green's function $G$
and the screened Coulomb interaction $W$. This implies
that screening is described within the theory, instead
of being introduced into it through an effective parameter.
Besides the screened Coulomb interaction $W$, the GW+DMFT 
theory introduces an effective local interaction $\mathcal{U}$
used as the bare interaction within an effective local model.
The GW+DMFT equations require
this interaction to be calculated self-consistently such as
to reproduce the local part of the fully screened interaction
$W$ when the local model is solved by many-body techniques.
This implies that the two interactions are related by a
two-particle Dyson (or Bethe-Salpeter) equation
$\mathcal{U}^{-1} - W_{loc}^{-1} =P_{loc}$, where $P_{loc}$
is the polarisation function of the local problem.
The physical content of this construction can be described
as follows: instead of using the full long-range Coulomb 
interaction within a full continuum description, an effective 
local interaction $\mathcal{U}$ is used within an effective local problem, 
but the interaction $\mathcal{U}$ is determined such that the two problems 
reproduce the same {\it fully screened local interaction} $W$.
The price to pay is that the effective interaction $\mathcal{U}$
inherits from the fully screened interaction $W$ its dynamical,
i.e. frequency-dependent character (even though the bare interaction
in the full Hilbert space, the bare Coulomb interaction, is 
frequency-independent).

Interestingly, this concept can be generalized and has
proven useful even outside the GW+DMFT scheme. Namely, the full
many-body problem can be simplified by eliminating some of the
interacting degrees of freedom, at the price of introducing
an effective dynamical interaction. The latter is determined
from the requirement that when the resulting many-body problem
is solved, the fully screened interaction is retrieved.
In this work, we describe the different sources of
frequency-dependence of effective local Hubbard interaction
and investigate their effects on solid state spectroscopies.
Section 2 discusses the dynamical character of effective
interactions, while section 3 presents the relation to coupled
electron-boson Hamiltonians. Solving such a Hamiltonian
within DMFT, that is, by mapping onto a local problem,
consists in generalizing the usual DMFT concept to a ``doubly
dynamical'' one, where not only the Weiss mean field is
dynamical but also the effective local interactions.
We abbreviate this ``doubly dynamical mean field theory''
in the following as ``DDMFT''.
Section 4 introduces approximate but very efficient and
accurate concepts for solving 
the dynamical impurity model arising within DDMFT
in the antiadiabatic limit. Section 5 addresses implications
for the resulting spectral functions, in particular with
respect to satellite structures and spectral weight transfers
-- concepts that are then applied to the ternary transition
metal oxide SrVO$_3$ in section 6.
A discussion of observable consequences of dynamical screening
effects concludes this work.

\section{Dynamical interactions: the concept of partial screening}

The above equation for the effective local interaction
$\mathcal{U}$ can be rewritten as
\begin{eqnarray} 
W_{loc} = \frac{\mathcal{U}}{1-P_{loc} \ \mathcal{U}}
\end{eqnarray}
stressing the interpretation of screening of the
effective interaction $\mathcal{U}$ by the 
dielectric function of the effective local problem:
\begin{eqnarray} 
\epsilon_{loc}^{-1} = \frac{1}{1 - P_{loc} \ \mathcal{U}}.
\end{eqnarray}
Alternatively, one can say that the screened interaction
is ``unscreened'' by $P_{loc}$ to obtain $\mathcal{U}$:
\begin{eqnarray} 
\mathcal{U} = \frac{W_{loc}}{1+P_{loc}W_{loc}}
\end{eqnarray}

These observations have motivated generalizations of the
concept of partial screening, where a many-body problem
is solved in a two-step procedure: first, an effective
Hamiltonian (or action) is constructed in a Hilbert space
that is a subset of the original space. Finally, this
effective many-body problem is solved with some suitable
many-body technique.
The bare interaction in the subspace is a partially 
screened interaction in the full space. In order to
determine it, one needs some estimates for the
fully screened interaction $W$ and the polarization
``at the second step'', the polarization
$P_{\textrm{step}-2}$ of the effective many-body problem.
Then, the effective interaction is constructed as
\begin{eqnarray} 
\mathcal{U} = \frac{W}{1+P_{\textrm{step}-2}W}.
\end{eqnarray}
The most important example of such a ``constrained
screening approach'' (see \cite{PhysRevB.86.165105} for a
more detailed discussion of the general philosophy) is 
the so-called ``constrained random phase approximation''
\cite{PhysRevB.70.195104}.
The cRPA provides an (approximate) answer to the following
question: given the Coulomb Hamiltonian in a large Hilbert
space, and a low-energy Hilbert space that is a subspace of
the former, what is the effective {\it bare} interaction to
be used in many-body calculations dealing only with the
low-energy subspace, in order for physical predictions for
the low-energy Hilbert space to be the same for the two
descriptions?
A general answer to this question not requiring much less
than a full solution of the initial many-body problem, the
cRPA builds on two approximations: it assumes (i) that the
requirement of the same physical predictions be fulfilled
as soon as in both cases the same estimate for the fully
screened Coulomb interaction $W$ is obtained
and (ii) the validity of the random phase approximation
to calculate this latter quantity.

The cRPA starts from a decomposition of the polarisation
of the solid in high- and low-energy parts, where the
latter is defined as given by all screening processes
that are confined to the low-energy subspace. The former
results from all remaining screening processes:
\begin{eqnarray} 
  P^{\rm high}=P - P^{\rm low},
\label{polsep}
\end{eqnarray}
One then calculates a partially screened interaction
\begin{eqnarray} \label{wrest}
  W^{\rm partial} =
\varepsilon_{\rm partial}^{-1} v. 
\end{eqnarray}
using the {\it partial} dielectric function
\begin{eqnarray} \label{dielectric}
  \varepsilon_{\rm partial}^{-1} = \frac{1}{1-
P^{\rm high}v}. 
\end{eqnarray}

Screening $W^{\rm partial}$ by processes that live within the 
low-energy space recovers the fully screened interaction $W$.
This justifies the interpretation of the matrix elements of 
$W^{\rm partial}$ in a localized Wannier basis as the interaction 
matrices to be used as bare Hubbard interactions within a low-energy
effective Hubbard-like Hamiltonian written in that Wannier basis.

Hubbard interactions -- obtained as the static ($\omega=0$) limit 
of the onsite matrix element
$\langle |W^{\rm partial} | \rangle$ within cRPA -- have by now been
obtained for a variety of systems, ranging from transition
metals \cite{PhysRevB.70.195104} to oxides
\cite{miyake:085122,PhysRevB.74.125106,jmt_mno, PhysRevB.86.165105},
pnictides \cite{miyake2008, nakamura, miyake2010, imada2008}, or 
f-electron compounds \cite{jmt_cesf}, and several implementations
within different electronic structure codes and basis sets
have been done, e.g. within 
linearized muffin tin orbitals \cite{PhysRevB.70.195104}, maximally
localized Wannier functions \cite{miyake:085122, cRPA-friedrich, nakamura}, or
localised orbitals constructed from projected atomic
orbitals \cite{PhysRevB.86.165105}.
The implementation into the framework of the Wien2k package
\cite{PhysRevB.86.165105} made it possible that Hubbard $U$'s be calculated
for the same orbitals as the ones used in subsequent LDA+DMFT 
calculations, see e.g.~\cite{martins, PRX-BaCoAs, BaTiFeAsO, Razzoli, CaFeAs}.

Still, the most important insight from the cRPA is probably the
fact that the effective low-energy Coulomb interactions are now
frequency-dependent. In complete analogy to the GW+DMFT equations
that provide the effective local $\mathcal{U}$ with a dynamical
character, the elimination of certain degrees of freedom leads
to an effective frequency dependence. Within GW+DMFT the downfolded
degrees of freedom are {\it nonlocal} processes resulting from
nonlocal interactions and polarisations, while the cRPA gives a
recipe for downfolding higher energy degrees of freedom.
An obvious idea is then to use both concepts, and perform GW+DMFT
calculations within a low-energy subspace with bare interactions
determined from cRPA. This route has been explored successfully
in \cite{hansmann, jpcm12}.

\section{Electron-plasmon Hamiltonians from first principles}

The above discussion motivates studies of effective problems
with frequency-dependent interactions. Within the GW approximation 
the dynamical character is naturally taken into account when
expanding in $W$. Still, while the high-frequency behavior
of $W$ calculated from the random phase approximation is often
a good approximation to the true one, exhibiting in particular
satellite structures at the plasma energy of the system, its
use in the first order formula $\Sigma = GW$ leads to artefacts:
when used to recalculate the Green's function (and from it, the
spectral function) the use of the GW self-energy truncates
the series of plasmon replicae to a single one, which is moreover
somewhat displaced in frequency \cite{0953-8984-11-42-201}.
This problem is cured when a cumulant form is used instead of
the GW self-energy \cite{arya-cumulant}, and much recent effort
has been spent to work out high-energy spectral functions within
a GW-based cumulant approach for different materials \cite{guzzo,
gatti, kas, nakamura-arXiv:1511.00218}.

Interestingly, recent work within a generalized DMFT context has 
allowed to bridge between the pictures of many-body perturbation
theory and lattice models for correlated fermions. Indeed, when
performing ``realistic'' DMFT calculations (that is, DMFT
calculations based on an Hamiltonian that is
extracted from first principles calculations) it has become 
possible by now to include the full frequency-dependence of
the effective local Hubbard interactions, and at high energies
the structure of the GW-based cumulant expansion is recovered.
At low energy, the DMFT-based picture leads to a generalization
of the cumulant approach as formulated in \cite{arya-cumulant}, since
the starting Green's function is itself an {\it interacting}
Green's function. We will come back to this point below.

Extending the philosophy of realistic DMFT calculations to dynamically
screened interactions requires the use of a framework that allows
for a description of an explicit frequency-dependence of the interactions
$\mathcal{U}(\omega)$.
One possibility is to switch from the Hamiltonian
formulation to an action description where
the frequency-dependent nature of the interaction is readily
incorporated as a retardation in the interaction term  
\begin{eqnarray}
S_{int}[\mathcal{U}] 
= - \int_0^{\beta} \int_0^{\beta} 
d\tau d\tau^{\prime} \mathcal{U}(\tau- \tau^{\prime})
n(\tau) n(\tau^{\prime})
\end{eqnarray}
where we have assumed that the retarded interaction couples only
to the density $n(\tau)$.
Alternatively, it is possible to stick to a Hamiltonian formulation.
In order to describe the retardation effects in the interaction one
then needs to introduce additional bosonic degrees of freedom that
parametrise the frequency-dependence of the interaction.
Indeed, from a physical point of view, screening can be understood
as a coupling of the electrons to bosonic screening degrees of freedom
such as particle-hole excitations, plasmons or more
complicated composite excitations giving rise to shake-up satellites
or similar features in spectroscopic probes.
Mathematically, a local retarded interaction can be represented
by a set of bosonic modes of frequencies $\omega$ coupling to the 
electronic density with strength $\lambda_{\omega}$.
The total Hamiltonian 
\begin{eqnarray}
H= H_{0} + H_V + H_{screening}
\end{eqnarray}
is then composed by a one-body part (e.g. of ``LDA++'' \cite{Lichtenstein98}
or screened exchange \cite{BaCoAs, vanroeke-epl, CaFeAs} form), 
a local interaction term $H_V$ that is of
Hubbard form
but with the local interactions given by the {\it unscreened}
local matrix elements of the bare Coulomb interactions $V$ and the
Hund's exchange coupling $J$ (assumed not to be screened by the
bosons and thus frequency-independent)
\begin{eqnarray}
H_V &=& 
\frac 12\sum_{imm^{\prime}\sigma } 
V_{mm^{\prime }}^in_{im\sigma}n_{im^{\prime }-\sigma }  
+ \frac 12\sum_{im\neq m^{\prime}\sigma} 
(V_{mm^{\prime }}^i-J_{mm^{\prime}}^i)n_{im\sigma }n_{im^{\prime }\sigma }  
\end{eqnarray}
and a screening part consisting of the local bosonic modes and their
coupling to the electronic density:
\begin{eqnarray}
H_{screening} = 
\sum_{i}\int d\omega \Big[\lambda_{i \omega}(b_{i\omega}^{\dagger} + b_{i\omega}) 
\sum_{m \sigma} n_{i m \sigma}+\omega \ b_{i\omega}^{\dagger}b_{i\omega}\Big].
\nonumber
\end{eqnarray}
As in standard LDA+DMFT, many-body interactions 
are included for a selected set of local orbitals, assumed
to be ``correlated''. The sums thus run over atomic sites $i$
and correlated orbitals $m$ centered on these sites.
In the first attempts putting up a ``LDA+$\mathcal{U}(\omega)$+DMFT''
scheme \cite{PhysRevB.85.035115, werner_Ba122}, $H_0$ is given by the Kohn-Sham 
Hamiltonian of DFT, suitably corrected for double counting terms. 
More recently, in the so-called ``screened exchange dynamical
mean field theory'' a screened exchange Hamiltonian is used as
a starting point \cite{vanroeke-epl,BaCoAs,CaFeAs}.

Integrating out the bosonic degrees of freedom would lead back
to a purely fermionic action with retarded local interactions
\begin{eqnarray}
\mathcal{U}(\omega) = V + \int d\omega^{\prime} 
\lambda_{\omega^{\prime}}^2   
\left(\frac{1}{\omega- \omega^{\prime}}
-\frac{1}{\omega+ \omega^{\prime}} \right)
\end{eqnarray}
The above Hamiltonian thus yields a parametrisation of the
problem with frequency-dependent interactions provided that
the parameters are chosen as
$\text{Im} \mathcal{U}(\omega)= - \pi\lambda_\omega^2$.
The zero-frequency (screened) limit is then given by
$U_0=V-2\int d\omega \frac{\lambda_{\omega}^2}{\omega} $.

\section{The Dynamic Atomic Limit Approximation}

In practice, an extremely efficient scheme for the solution of this
problem, suitable in the antiadiabatic regime, can be obtained
within a dynamical mean field framework, when the DMFT equations
are solved by the recently introduced \cite{PhysRevB.85.035115} ``Boson
factor ansatz'' (BFA). As shown in \cite{krivenko},
this scheme can in fact be understood as the zeroth order
(in the hybridization) approximation to a set of slave rotor
equations.
Dynamical mean field theory maps the lattice problem (or here,
the solid) onto an effective local (``impurity'') problem.
The new aspect in the present context is the dynamical character
of the interaction in this local impurity problem.
The BFA consists in approximating the local 
Green's function of the dynamical impurity model as
follows:
\begin{equation}
G(\tau )= - \langle \mathcal{T} c(\tau) c^{\dagger}(0) \rangle
=
\left( \frac{G(\tau )}{G_{stat}(\tau )}\right) 
G_{stat}(\tau )
\sim
\left( \frac{G(\tau )}{G_{stat}(\tau )}\right) {\bigg|}_{\Delta=0}
G_{stat}(\tau )
\label{factorisation-approximation}
\end{equation}%
where G$_{stat}$ is the Green's function of a fully interacting
impurity model with {\it purely static interaction} U=$%
U(\omega =0)$, and the first factor is approximated by its
value for vanishing bath hybridization $\Delta$ \cite{PhysRevB.85.035115}.
In this case, it can be analytically evaluated in terms of 
the frequency-dependent interaction:
\begin{equation}
B(\tau) =
\left( \frac{G(\tau )}{G_{stat}(\tau )} \right) {\bigg|}_{\Delta=0}
= e^{- \int_0^{\infty} \frac{d\omega}{\pi}
\frac{|\mathrm{Im} U(\omega)|}{\omega^2}( K_{\tau}(\omega) - K_{0}(\omega))}
\label{factorisation-approximation} 
\end{equation}%
with $K_{\tau}(\omega)=\frac{exp(-\tau \omega) + exp(-(\beta- \tau) \omega)}
{1 - exp(-\beta \omega)}$.
In the regime that we are interested in, namely 
when the plasma frequency that characterises the variation of
$\mathcal{U}$ from the partially screened to the bare value, is
typically several times the bandwidth, this is an excellent approximation, as
was checked by benchmarks against direct Monte Carlo calculations in
Ref.~\cite{PhysRevB.85.035115}. 
The reason can be understood when considering the solution of the
dynamical local model in the {\it dynamical atomic limit}
$\Delta=0$, that is,
when there are no hopping processes possible between the impurity
site and the bath. In this case the BFA trivially yields the exact
the solution, and the factorisation can be understood as a factorisation
into a Green's function determined by the static Fourier component
of $\mathcal{U}$ only and the exponential factor $B$ which only depends
on the non-zero frequency components of $\mathcal{U}$.
The former fully determines the low-energy spectral function of the
problem, while the latter is responsible for generating high-energy
replicae of the low-energy spectrum.
For finite bath hybridisation, the approximation consists 
in assuming that the factorisation still holds and that the
finite bath hybridisation modifies only the low-energy
static-$U$ Green's function, leaving the general structure of the
plasmon replicae generation untouched. The approximation thus
relies on the energy scale separation between low-energy processes
and plasmon energy; it becomes trivially exact not only in the
atomic limit but also in the static limit, given by small
electron-boson couplings or large plasmon energy.

Interestingly, the scheme is strongly reminiscent of cumulant approaches
derived from the GW approximation \cite{arya-cumulant}. The main
difference -- except for the restriction to the local picture in the
present formulation -- is the fact that the prefactor of the
cumulant exponential is itself a many-body Green's function that
cannot in general be represented by a non-interacting band structure.
It is, however, restricted to a purely local
description of satellite features and as such not a good
approximation e.g. to plasmon dispersions.
Technically, a difference appears also through the use of the
cRPA interaction $\mathcal{U}$ instead of the fully screened
interaction. Conceptually, in the spirit of the GW+DMFT scheme
one would eventually like to use a partially screened interaction
that is not only screened  by high-energy degrees of freedom
(as done here) but also by nonlocal screening processes in the
sense of GW+DMFT.

\section{Satellites and Spectral Weight}

The BFA lends itself naturally to a mathematical formulation 
of the generation of plasmon replicae. Indeed, the
factorisation of the Green's function corresponds in frequency
space to a convolution of the spectral representations of the
low-energy Green's function $G_{static}$ and the bosonic factor $B$.
In terms of the spectral
function $A_{stat}(\omega)$ of the static Green's function $G_{stat}(\omega)$
and the (bosonic) spectral function $\rho_B(\epsilon)$ of the bosonic
factor $B(\tau)$ defined above
the spectral function $A(\omega)$ of the full Green's function
$G(\tau)$ reads:
\begin{equation}
A(\omega)=\int_{-\infty}^\infty \!\!\! d\epsilon ~ \rho_B(\epsilon) 
\frac{1+e^{-\beta\omega}}{(1 + e^{-\beta(\epsilon-\omega)})(1 - e^{-\beta\epsilon})} 
A_{stat}(\omega-\epsilon).
\label{spectral_conv}
\end{equation}
In the case of a single mode of frequency $\omega_0$, the bosonic
spectral function consists of sharp peaks at energies given by
that frequency, and the convolution generates replicae of the
spectral function $A_{stat}(\omega)$ of the static part.
Due to the overall normalisation of the spectral function,
the appearance of replicae satellites is necessarily accompanied
by a transfer of spectral weight to high-energies. This 
mechanism induces a corresponding loss of spectral weight in the
low-energy part of the spectral function. Indeed, it can be
shown \cite{casula_effmodel}
that the spectral weight corresponding to the low-energy
part as defined by a projection on zero boson states is 
reduced by the factor
\begin{eqnarray}
Z_B & = & \exp \left( - 1/\pi  \int_0^\infty \!\!\! d \nu 
  ~  |\textrm{Im} \mathcal{U}(\nu)| /\nu^2 \right).
\label{Z_B}
\end{eqnarray}
Estimates of $Z_B$ for typical transition metal oxides vary
between $0.5$ and $0.9$, depending on the energy scale of
the plasma frequency and the efficiency of screening
(as measured e.g. by the difference between bare Coulomb
interaction 
$\langle |\frac{1}{|r - r^{\prime}|}| \rangle = \mathcal{U}(\omega=\infty)$
and the static value $\mathcal{U}(\omega=0)$).

\section{Illustration on the example of SrVO$_3$}

\begin{figure}%
\
\\
\
\\
\
\\
\
\\
\
\\
\
\\
\
\\
\
\\
\includegraphics[width=10cm]{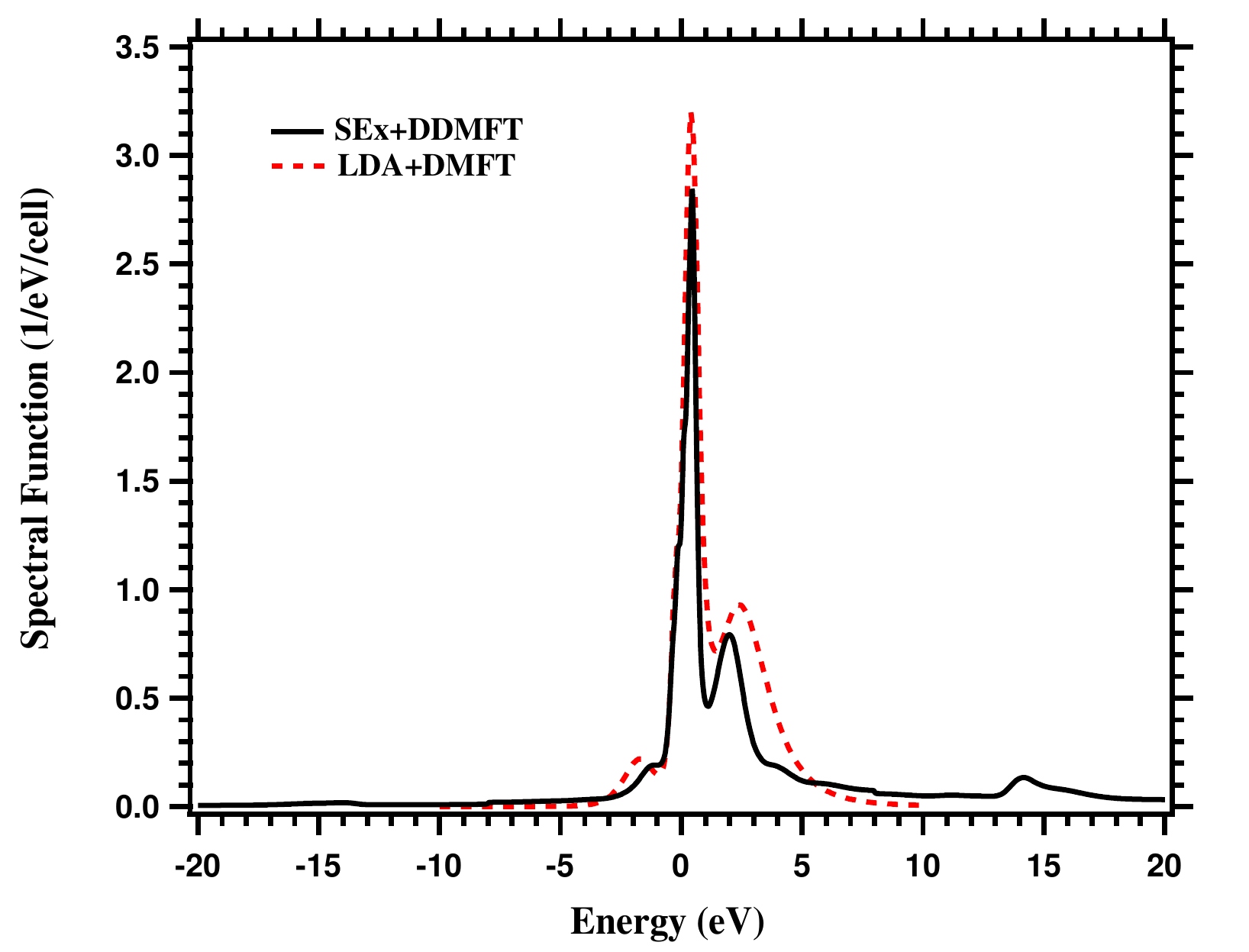}
\caption{Spectral function of a low-energy (t$_{2g}$-only)
screened exchange Hamiltonian within standard LDA+DMFT 
with U=4 eV and J=0.6 eV (red
curve) and ``Screened Exchange + DDMFT''
(``SEx+DDMFT'') (black curve). The calculation uses the same setup 
as in \cite{vanroeke-epl}. Besides the low-energy electronic structure
with lower and upper Hubbard bands (at -1.5 eV and 2.0 eV
respectively) and a renormalized quasi-particle peak at the
Fermi level, the plasmon satellites at $\pm$ 15 eV are clearly
visible when the dynamical character of the interaction is
included as done in SEx+DDMFT. 
}%
\label{svo-dynU}%
\end{figure}

To illustrate the effects of the dynamical interactions, we
present in Figs.~1 and 4 the spectral function of the t$_{2g}$
states of the ternary transition metal oxide SrVO$_3$ within 
different schemes. SrVO$_3$ is a perfectly cubic perovskite
with d$^1$ configuration. The crystal field splitting between
e$_g$ and t$_{2g}$ states suggests a many-body description in
terms of the t$_{2g}$ states only, and this was done in the
present case.
It should be noted however that the contribution of unoccupied
e$_g$ states dominates at energies as low as $\sim 2.5$ eV.
This issue has been discussed in detail in \cite{jmt_svo2}.

The compound is a correlated metal with a moderate quasi-particle
renormalisation, and has been the subject of experimental
(see e.g. \cite{maiti01,
PhysRevB.73.052508, PhysRevLett.93.156402,
PhysRevB.80.235104, PhysRevB.52.13711} and references thereof)
and theoretical (see e.g. \cite{Ishida10, pavarini:176403, 
mossanek:033104}) 
studies over the last 20 years.
Detailed spectroscopic, transport, and thermodynamic data
are available, and innumerous theoretical works have 
used this compound as a benchmark compound for new calculational
methods \cite{PhysRevB.87.115110, nomura}. 
Among others, the first calculations within DDMFT
\cite{casula_effmodel} and a dynamical implementation of the
combined GW+DMFT scheme \cite{jmt_svo, jmt_svo2} have been performed
on SrVO$_3$.
An overview with the respective references can be found
in \cite{jmt_svo, jmt_svo2}.

\begin{figure}%
\
\\
\
\\
\
\\
\
\\
\
\\
\
\\
\
\\
\
\\
\includegraphics[width=10cm]{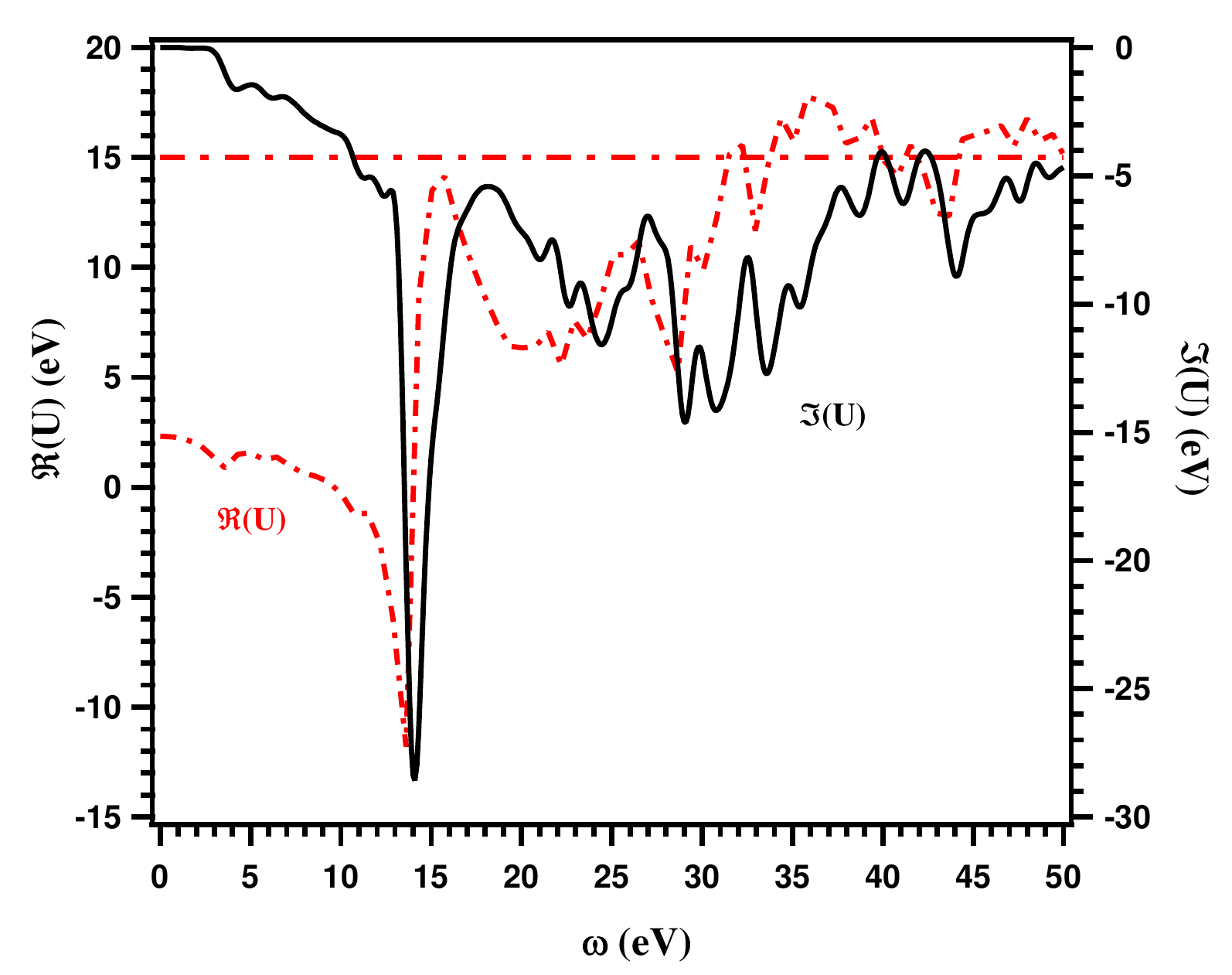}
\caption{
Dynamical interaction for  a t$_{2g}$-only Hamiltonian
for SrVO$_3$ calculated within the cRPA approximation.
}%
\label{svo-dynU}%
\end{figure}

Figure 1 displays a calculation for SrVO$_3$ within
DDMFT with the dynamical interaction calculated from the cRPA.
Here, a screened exchange Hamiltonian was used as the one-body
part of the Hamiltonian, following the proposal of ``Screened
Exchange Dynamical Mean Field Theory'' of Refs. \cite{vanroeke-epl,BaCoAs}.
Technical details of the calculation can be found in \cite{vanroeke-epl}.
The t$_{2g}$ states present at the Fermi level are narrowed into
a thin quasi-particle peak, and lower and upper Hubbard bands 
are visible at -1.5 eV and 2.0 eV respectively.
The dynamical interactions lead moreover to an additional
transfert of spectral weight from the low-energy part of the
spectrum to high energy (plasmon-) satellites, appearing at
$\pm$ 15 eV. They correspond to photoemission or inverse
photoemission processes that imply the creation or destruction
of a bosonic excitation of 15 eV. 
Their energy is compatible with the plasmon
features measured at multiples of 15 eV within low-energy
electron diffraction measurements for the related SrTiO$_3$
\cite{eels}.

\begin{figure}%
\
\\
\
\\
\
\\
\
\\
\
\\
\
\\
\
\\
\
\\
\includegraphics[width=10cm]{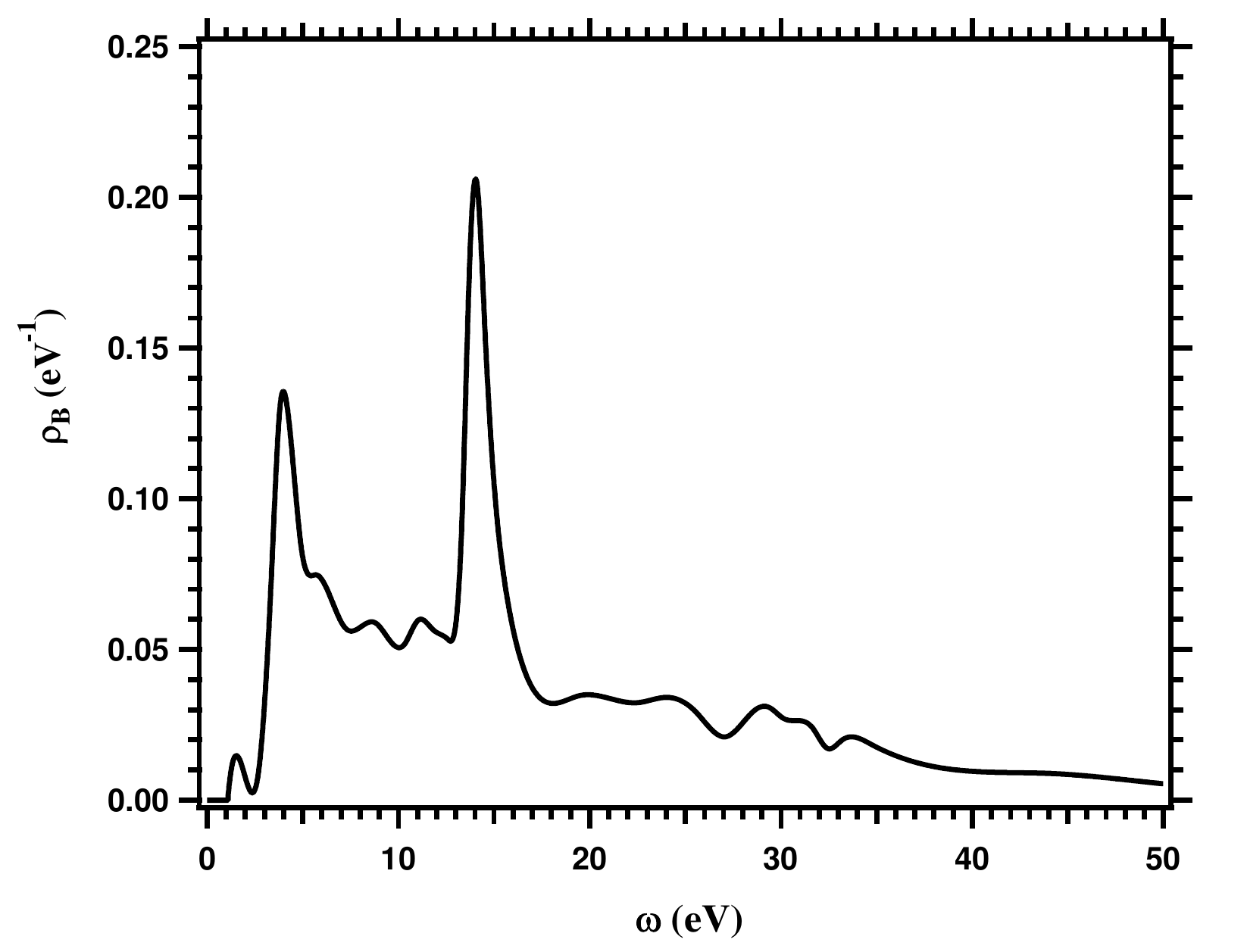}
\caption{Spectral function $\rho_B$ of the bosonic factor $B(\tau)$
for SrVO$_3$. The strong peak at 15 eV corresponds to the
plasmon seen in $\mathcal{U}$. A second ``sub-plasmon''
is visible around 5 eV. This feature is more clearly seen
in the fully screened interaction $W$, since at this energy
there is a collective plasmon corresponding to the t$_{2g}$
charge only. Since transitions within this manifold are
cut out when calculating $\mathcal{U}$, only remnants
of this feature (surviving due to hybridisation effects)
are left within $\mathcal{U}$.
Convolution of this curve with the spectral function corresponding
to a calculation with static interactions generates the final
spectral function. The pronounced peak structures can therefore
be directly related to satellite features in the final spectra.
}%
\label{svo-rhoB}%
\end{figure}

The most interesting aspect in the light of the present discussion
is the appearance of the high-energy satellites structures, along 
with the corresponding spectral weight transfert. Since the
overall spectral function is normalised, the appearance of 
satellites necessarily reduces the spectral weight in the
low-energy part of the spectrum.
The distribution of spectral weight between the low-energy
part (-2 eV to 2eV) around the Fermi level, and the bosonic
satellite structures corresponds to the values discussed
above: the imaginary part of $\mathcal{U}$ (Figure 2)
leads to a 
$Z_B$ factor (Eq. (15)) of 0.6, corresponding to a ratio
of 0.6:0.4 for the
low-energy spectral weight to the weight of the
high-energy satellites.

These spectral properties are thus the direct consequence of
the dynamical interaction, plotted in Figure 2.
As a partially screened interaction $\mathcal{U}$ is
related to a (partial) charge-charge response function, and
as such has real and imaginary parts (red and black lines
respectively) related by a Kramers-Kronig relation.
The real part is characterized by its low-energy value of about
3 eV, corresponding to the usual static Hubbard interaction
and its high-energy tail recovering the unscreened interaction
in the infinite frequency limit. The change of regime happens
at the plasma frequency of about 15 eV. The imaginary part can
be understood as the density of screening modes (plasmons,
particle-hole excitations ...). The sharp peak at 15 eV corresponds
to the plasmon that is responsible for the pronounced structure
of the real part.

To illustrate more clearly the consequences of the dynamical
interaction, we present in Figure 4 a series of calculations
where we have a started from a standard LDA+DMFT calculation
with static interactions (blue curve) 
and used this curve as an
approximation to the
spectral function $A_{stat}$ in Eq. (14).
The black curve 
is obtained by performing the convolution with the
bosonic spectral weight function $\rho_B$ of Eq. (15),
see Figure 3 and the Appendix for details.
Besides the case of the physical $\rho_B$, we show two cases
where the electron-boson coupling has been artificially
enhanced. This is done by simply multiplying the spectral
distribution of modes $\Im \mathcal{U}$ by multiplicative 
factors (red and greens curves).

\begin{figure}%
\
\\
\
\\
\
\\
\
\\
\
\\
\
\\
\includegraphics[width=10cm]{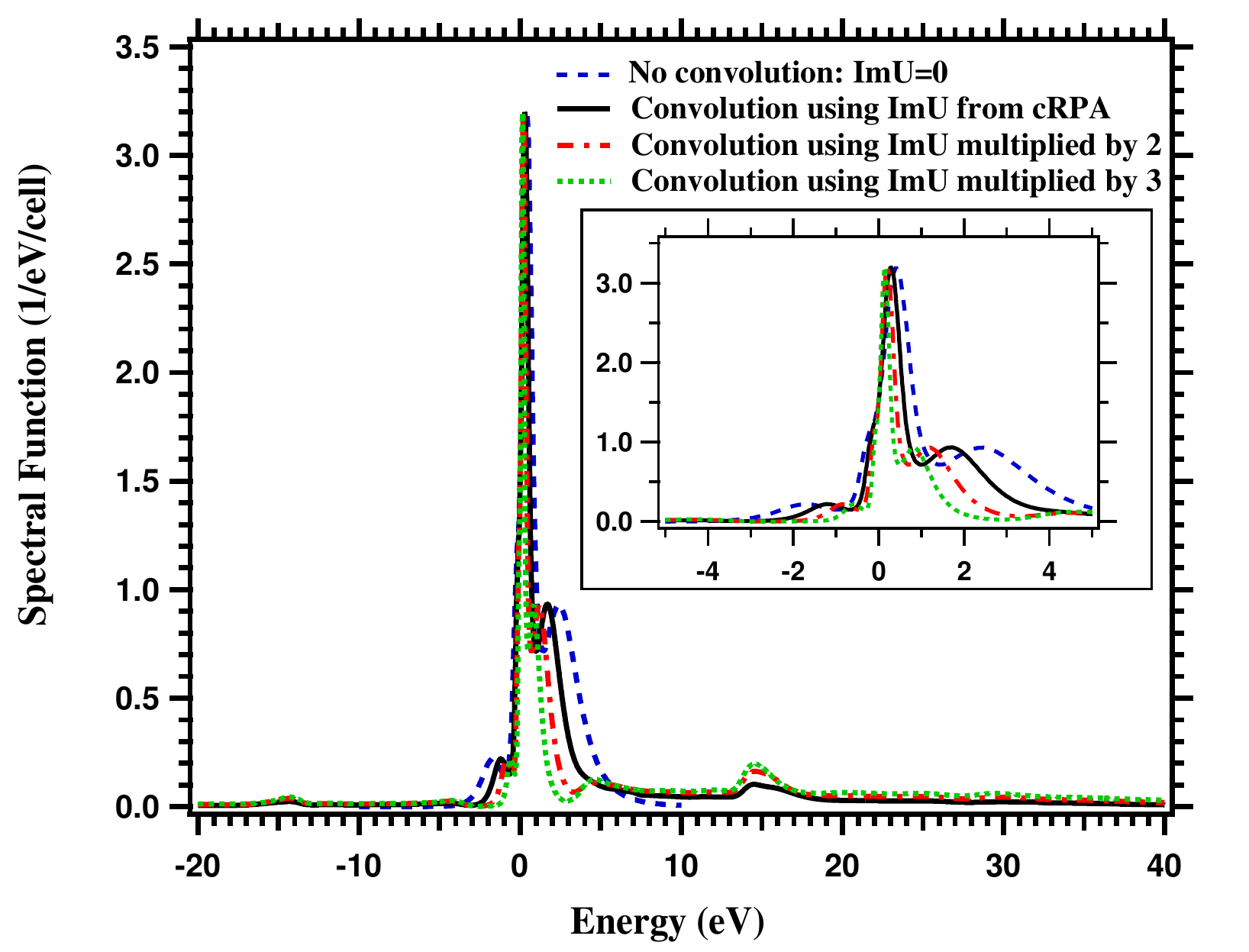}
\caption{Spectral function of a low-energy (t$_{2g}$-only) Hamiltonian
produced within a simple model construction:
we use an LDA+DMFT spectral function calculated for a static
interaction of U=4 eV, J=0.6 eV as approximation to the
spectral function $A_{aux}$ in Eq. (14), and obtain the final
spectral function by performing the convolution with the
bosonic spectral weight function $\rho_B$ of Eq. (15).
Besides the case of the physical $\rho_B$, we show two cases
where the electron-boson coupling has been artificially
enhanced. This is done by simply multiplying the spectral
distribution of modes $\Im \mathcal{U}$ by multiplicative
factors. 
}%
\label{svo-dynU}%
\end{figure}

The figure nicely demonstrates the increasing spectral
weight transfert when the electron-boson coupling becomes
larger. Now, even the second plasmon satellite structure
at 30 eV, corresponding to the creation of two 15 eV
plasmons can be seen. Also, smaller satellite features, 
corresponding to a sub-plasmon at about 5 eV that carry too little
spectral weight to clearly appear in the physical case now
become visible.
The spectral weight transfert to higher energy not only
reduces the low-energy spectral weight, but also leads
to a corresponding reduction of the energy scales in the
low-energy zero plasmon part of the spectrum. This is
expected: indeed, the mechanism of reduction is to rescale
all hoppings with $Z_B$. For a mathematical derivation of
this relation, see \cite{casula_effmodel}. Physically it
can be understood by noting that the basis
that diagonalizes the electron-boson Hamiltonian corresponds
to polaronic states, electrons dressed by their respective
screening clouds. This ``electronic polaron'' effect
\cite{biermann-JPCM, Macridin, 0953-8984-11-42-201}
effectively enhances the masses of the charge carriers
and translates itself into a narrowed band width.

\section{Conclusions and perspectives: 
What measurable consequences to the physics of dynamical
screening?}

We close this work with a discussion of measurable
consequences of the above phenomena. Obviously, a first
indicator for the physics discussed above are satellite
structures. Indeed, plasmonic satellites are ubiquitous
in photoemission spectroscopy and have -- in the past
-- been considered rather as a nuisance. A systematic
investigation, however, could provide extremely useful
information about the strength of Coulomb interactions
and screening as well as mobile charge carriers.
An effect that is harder to observe is the reduction
of spectral weight, since photoemission spectroscopy
does not have access to absolute values of the spectral
function. Probes that assess spectral weight as absolute
values, however, such as optical measurements, should
be sensitive to this kind of effects.
As discussed above, it should also be possible to
diagnose the spectral weight reduction as a reduction
in hoppings (and therefore bandwidth). This raises
however the question of the suitable starting band
structure. Indeed, as discussed in \cite{BaCoAs},
there is no a priori reason to use the Kohn-Sham
band structure of DFT as input to a many-body 
calculation. If a generalized one-body Hamiltonian,
based on screened exchange, is employed there is
a large compensation effect between band widening
by screened exchange and the band narrowing by
$Z_B$. As far as only questions of overall bandwidth
are concerned, a many-body calculation with static
interaction and using a DFT Hamiltonian is then
in fact quite a good approximation. If the spectral
weight was assessible quantitatively from experiment,
serious discrepancies should however be observed.
Finally, we also note that the reshuffling of
states around the Fermi level is also expected
to impact the magnetic properties of compounds
with high densities of states. Ref. \cite{BaCoAs}
for example, explained the absence of ferromagnetism
in BaCo$_2$As$_2$ by the rearrangement of electronic
states around the Fermi level when corrections due
to screened exchange and dynamical interactions
are included.

The related BaFe$_2$As$_2$ compound is an 
example where satellite features of the above
kind have been identified in photoemission spectra
(see the data in \cite{Ding08, Yi09} and the 
discussion in \cite{werner_Ba122}). 
Similar satellite features are also ubiquitous in
transition metal oxides 
\cite{nacoo, lanio, cacro}. 
Still, systematic
studies of such effects have at present not yet
been worked out and can be expected to bring additional
valuable information concerning the response properties
of solids, the strength of electronic correlations 
and their first principles description.

\section{Acknowledgements}

This work would not have been possible without the series
of recent works on downfolding techniques for many-body
interactions, GW+DMFT and dynamical screening
\cite{PhysRevB.70.195104, PhysRevB.86.165105, 
casula_effmodel, 
PhysRevB.85.035115, werner_Ba122, 
gwdmft, ayral, ayral-prb, 
jmt_svo, jmt_svo2, hansmann, krivenko, vanroeke-epl}. 
SB thanks her respective coauthors
F. Aryasetiawan, T. Ayral, M. Casula, M. Ferrero,
A. Georges, P. Hansmann, 
M. Imada, H. Jiang, I. Krivenko,
A. Millis, T. Miyake, O. Parcollet,
A. Rubtsov,
J.M. Tomczak, L. Vaugier, P. Werner
for the enjoyable collaborations and D. D. Sarma, V. Brouet,
H. Ding, A. Fujimori, K. Maiti, P. Richard and
T. Yoshida for stimulating discussions.
Our special thanks go to Ashish Chainani for his interest in 
this work, his careful reading of the manuscript and 
most useful suggestions.
\\
The work was further supported by the European Research Council
under project 617196 (CORRELMAT)
and IDRIS/GENCI under project t2015091393.

\section{Appendix}
In this Appendix, we comment more precisely on the way we proceed
to artificially enhance the plasmonic couplings, in order to plot
Figure 4. Indeed,
we make a transformation that allows us to recover the Lang-Firsov
limit at low-energy and the DALA approximation at high energy. Starting
from a spectral function $A$ without plasmonic interactions and a
bosonic spectral function $\rho_{B}$ that corresponds to a transfer
of weight $1-Z_{B}$ to plasmonic excitations, we first define 
$\tilde{A}^{aux}(\omega)=A(\omega/Z_{B})/Z_{B}$.
This spectral function is still normalized and corresponds to the
Lang-Firsov limit, where the effective mass is enhanced by $Z_{B}^{-1}$
without taking into account any spectral weight transfer. We then
calculate the final spectral function $\tilde{A}$ in the DALA form:
\[
\tilde{A}(\omega)=\int_{-\infty}^{+\infty}d\epsilon\tilde{A}^{aux}(\epsilon)
\rho_{B}(\omega-\epsilon)\left[n_{F}(\epsilon)+n_{B}(\epsilon-\omega)\right]
\]
where $\rho_{B}$ has a regular part $\rho_{B}^{regular}$ and a delta
peak of weight $Z_{B}$:
\[
\tilde{A}(\omega)=Z_{B}\tilde{A}^{aux}(\omega)+\int_{-\infty}^{+\infty}d\epsilon
\tilde{A}^{aux}(\epsilon)\rho_{B}^{regular}(\omega-\epsilon)\left[n_{F}(\epsilon)
+n_{B}(\epsilon-\omega)\right]
\].

In order to articially enhance the effect of the plasmons,
we multiply $\Im \mathcal{U}(\omega)$ by the respective factors.
This procedure corresponds to a uniform enhancement of the 
plasmon-electron coupling for all plasmon frequencies, and 
translates itself into a highly nonlinear modification of the
Boson spectral function $\rho_B$ entering the convolution above
\cite{PhysRevB.85.035115}.

\end{document}